\documentclass[
  aps,
  prapplied,
  twocolumn,
  superscriptaddress,
  nofootinbib,
  longbibliography
]{revtex4-2}
\usepackage{mathptmx}
\usepackage{subfig}
\usepackage{psfrag,graphicx}
\usepackage{dcolumn}
\usepackage{amsmath,amssymb}
\usepackage{bm}
\usepackage{color}
\usepackage{latexsym}
\usepackage{epstopdf}
\usepackage{color}
\usepackage[english]{babel}
\usepackage{latexsym}
\usepackage{psfrag,graphicx}
\usepackage{amssymb}
\usepackage{amsfonts}
\usepackage{bm}
\usepackage{epstopdf}
\usepackage{color}
\usepackage{amsmath}
\usepackage{appendix}
\usepackage{tipa, phonetic}
\usepackage{subcaption}
\usepackage{graphicx}
\usepackage{amsmath} 
\usepackage{braket}
\usepackage{caption}

\def\be{ \begin{equation}}
\def\ee{ \end{equation}}
\def\bse{  \begin{subequations}}
\def\ese{  \end{subequations}}
\newcommand{\bea}{\begin{align}}
\newcommand{\eea}{\end{align}}

\def\bi{\begin{itemize}}
\def\ei{\end{itemize}}

\def\bt{\begin{tabular}}
\def\et{\end{tabular}}

\def\3half{\tfrac32}
\def\be{\begin{equation}}
\def\ee{\end{equation}}
\def\bea{\begin{align}}
\def\eea{\end{align}}
\def\bi{\begin{itemize}}
\def\ei{\end{itemize}}

\begin{document}

\author{K. N. Zlatanov}
\affiliation{Center for Quantum Technologies, Department of Physics, Sofia University, James Bourchier 5 blvd., 1164 Sofia, Bulgaria}
\affiliation{Georgi Nadjakov Institute of Solid State Physics, Bulgarian Academy of Sciences, 72, Tzarigradsko Chaussee Blvd., 1784 Sofia, Bulgaria}
\author{M. Mallweger}
\affiliation{Department of Physics, Stockholm University, 10691 Stockholm, Sweden}
\author{M. Hennrich}
\affiliation{Department of Physics, Stockholm University, 10691 Stockholm, Sweden}
\author{N. V. Vitanov}
\affiliation{Center for Quantum Technologies, Department of Physics, Sofia University, James Bourchier 5 blvd., 1164 Sofia, Bulgaria}
\title{Decoupling of the STIRAP and Microwave-Dressing paths in Trapped Rydberg Ion Gates}
\date{\today }

\begin{abstract}
The strong dipole-dipole interaction of trapped Rydberg ions offers the possibility of sub-microsecond entanglement gates. For example a two-qubit Control-Phase gate in ${}^{88}\text{Sr}^+$ ions can be realized, by simultaneous excitation to the Rydberg states via stimulated Raman adiabatic passage (STIRAP) with simultaneous microwave induced dipole-dipole interaction. We show that this excitation protocol distorts the dark-state of the STIRAP stage and is prone to decay from the intermediate state. Here, we propose a novel pulse ordering, in which the STIRAP and the microwave dressing of the Rydberg states occurs in separate stages, preventing mutual interference effects that are detrimental to the gate fidelity.  We show that, for experimentally feasible parameters, the proposed excitation scheme can achieve a fidelity of $99.93\%,$ surpassing the experimentally demonstrated gate. In addition, we demonstrate a non-adiabatic speed-up to 400 ns by employing asymmetric pulse shapes in the STIRAP stage. The entangling phase is then controlled solely through the interaction strength by nonresonant asymmetric chirping of the microwave field.
\end{abstract}

\maketitle
\section{Introduction}
Trapped ion platforms are at the forefront of quantum technologies, operating as sensors~\cite{Ilias2022,Gilmore2021,Bonus2025}, simulators~\cite{Monroe2021,Manovitz2020} and quantum information devices~\cite{Colin2019, Pogorelov2021}. They have ultra-high single-qubit fidelity~\cite{Leu2023,Smith2024} and recent developments allow them to be hosted on microfabricated surface traps~\cite{Romaszko2020} that promise high scalability by allowing shuttling of the qubits to different interaction zones, where the quantum information can be processed. The native two-qubit gate, most often based on the Molmer-Sorensen interaction~\cite{Molmer1998,Molmer2000}, generates very high fidelity~\cite{Clark2021,Gaebler2016} and can be robust to multiple error sources simultaneously~\cite{Zlatanov2025}. Its speed, however remains in the microsecond time scale~\cite{Li2025}. Speeding up the two-qubit gate is not a trivial task which is rooted in two problems. One can either boost the laser power, which eventually generates a detrimental a.c. Stark shift or increase the Lamb-Dicke parameter and risk going beyond the linear phonon regime, where multiple phonon quanta are created (annihilated) and thus the approximation $\exp\{i\eta (a^{\dagger}e^{i \nu}+a e^{-i \nu})\}\approx 1+i\eta (a^{\dagger} e^{i \nu}+a e^{-i \nu})$ fails. Although Stark shifts can be compensated~\cite{Tu2025} and some recent proposals go a few steps beyond the Lamb-Dicke regime~\cite{Kirchhoff2025}, a speed up that brings the gate to the nanosecond regime using the conventional approach remains elusive. 

An alternative to the photon-phonon mediated interaction for two-qubit gates is an idea borrowed from the trapped-atom community~\cite{Urban2009,Gaetan2009,Wilk2010} based on Rydberg interaction between the atoms. Rydberg-mediated quantum gates~\cite{Saffman2010} rely on the strong interactions that arise when atoms are excited to high-lying electronic states. There are several distinct interaction mechanisms, spanning over the energy of the  Rydberg states, distance between the qubits, strength and detuning of the external fields, by which a gate can be realized.
At large detuning from any resonances, the interaction is of the van der Waals type~\cite{Tong2004,Singer2004,Beguin2013}, which falls off rapidly with distance but grows very strongly with the principal quantum number. This regime underlies the original blockade gate proposals~\cite{Jaksch2000,Lukin2001}, where the presence of one excited atom shifts the energy of a second excitation and thereby prevents it.
When two Rydberg states are nearly resonant with another pair of levels, the interaction is strongly enhanced through a process known as a Förster resonance~\cite{Vogt2006,Ryabtsev2010,Tretyakov2017,Ryabtsev2018,Cheinet2020}. In this case, the coupling is effectively resonant and much longer-ranged than the van der Waals interaction. Förster resonances can be tuned by Stark shifts~\cite{Nipper2012} or external electric field~\cite{Kondo2016}, providing a high degree of experimental control.
While van der Waals interactions and Förster resonances form the foundation of many neutral-atom Rydberg gate proposals, these mechanisms are not directly applicable in trapped ions. In ions the van der Waals forces are significantly weaker due to the charged ion core, also the energy scaling is different and the Rydberg states cannot be brought easily to a Förster resonance. 

In trapped ions~\cite{SchmidtKaler2011}, Rydberg excitation can be combined with Coulomb interactions, adding yet another route to implement entangling gate~\cite{Vogel2019} or quantum simulations~\cite{Martins2023}. The interaction between the qubits can be realized when the ions are prepared in states with induced dipole moments~\cite{Muller2008}. When microwave fields are applied, coupling two different Rydberg states, oscillating dipoles are created. This effect offers a strong and tunable interaction~\cite{Petrosyan2014,Tanasittikosol2011,Bohorquez2023}.
The resulting dipole-dipole interaction has attracted attention~\cite{SchmidtKaler2020,Higgins2017b} because it allows speeds of operation in the nanosecond time scale. 
For example in \cite{Zhang2020} a two-qubit Control-Phase gate has been shown experimentally with gate time of 700 ns and $78\%$ fidelity and theoretical proposals envision even faster gates~\cite{Wilkinson2025}.

Although single photon excitation of the Rydberg states has been demonstrated, it requires radiation with very short wavelength, for example 122 nm~\cite{Feldker2015}. Such laser sources are not commonly accessible, the radiation is hard to control and also brings additional complications for all other optical devices, since they need to have decent efficiency at that wavelength. Therefore a two-photon excitation is more feasible, although it is still uv light (243 nm and 305 nm) \cite{Higgins2017}. In order to suppress the decay from the intermediate state one can employ STIRAP. This setup was employed in~\cite{Zhang2020}, where the microwave field is constantly ON during the gate.  

In this paper we show that if the STIRAP and the dressing field act simultaneously this creates a four-level system and distorts the dark state formed by STIRAP. We propose an excitation scheme which decouples the STIRAP and the dressing stages and improves its fidelity, beyond what is demonstrated in \cite{Zhang2020} experimentally. By using asymmetric pulse shapes we speed up the STIRAP stage beyond the adiabatic approximation, which is crucial for the overall gate speed. Within the dressing stage we realize complete population return (CPR), that allows the initialized Rydberg state to accumulate an entangling phase. We demonstrate that by employing non-resonant asymmetric chirping of the microwave field the phase control can be transferred to the dipole-dipole interaction alone. We show that this way one can attain $99.93\%$ fidelity assuming experimentally feasible parameters.

This paper is organized as follows, in Section \ref{sec:Two} we give a short introduction to the problem and we illustrate the interference between the single-qubit STIRAP with the dressing field. In Section \ref{sec:Three} we discuss the speeding up of the gate by asymmetric pulses based on the Dykhne-Davis-Pechukas(DDP) approximation. In Section \ref{sec:Four} we demonstrate the two-qubit gate with pulse shaping and chirping effects and we conclude in Section \ref{sec:Five}.

\section{DISTORTED STIRAP}\label{sec:Two}

\begin{figure}[t]
\bt{c}
 \includegraphics[width=0.99\columnwidth]{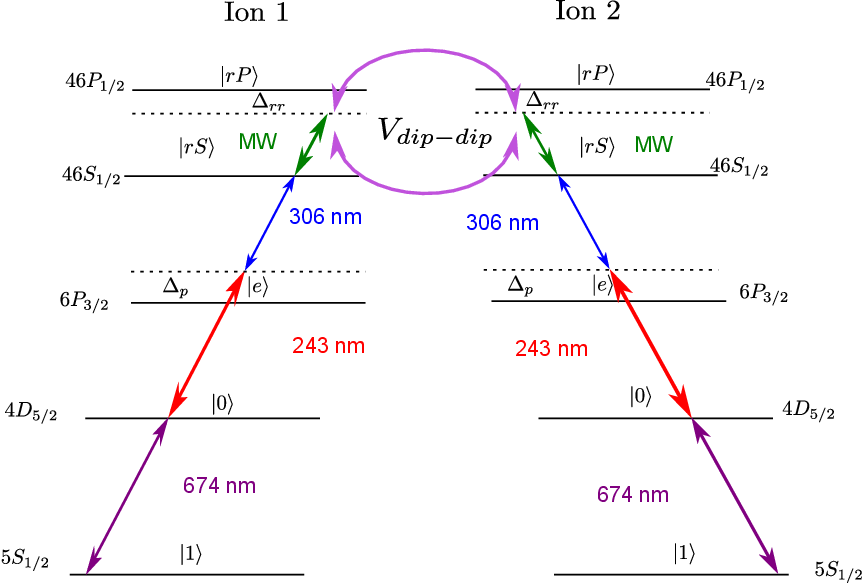}
\et
\caption{(Color online)
Energy configuration of the levels of the two ${}^{88}Sr^{+}$ ions participating in the Rydberg excitation. }\label{fig:1}
\end{figure}

\begin{figure*}[t]
  \noindent\makebox[\textwidth]{%
    \begin{tabular}{cc}
      \includegraphics[width=0.5\textwidth]{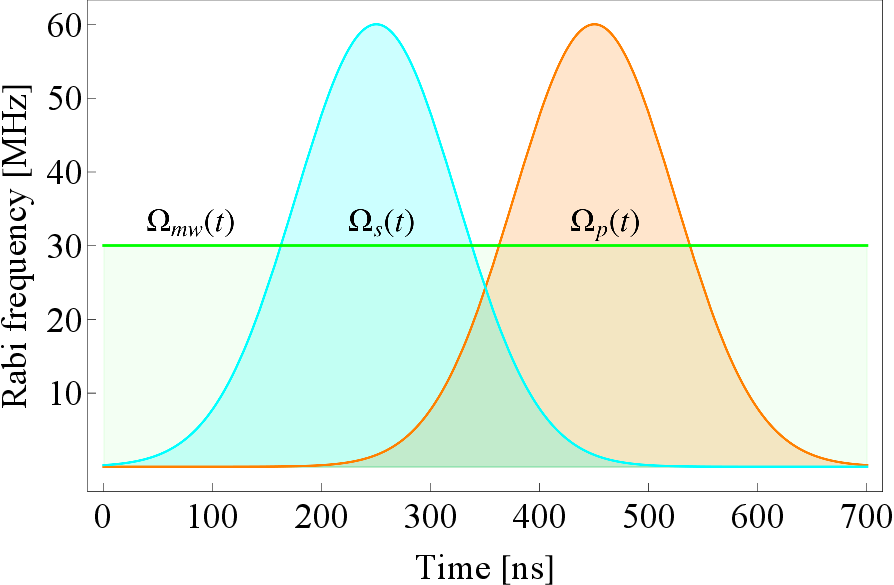} &
      \includegraphics[width=0.5\textwidth]{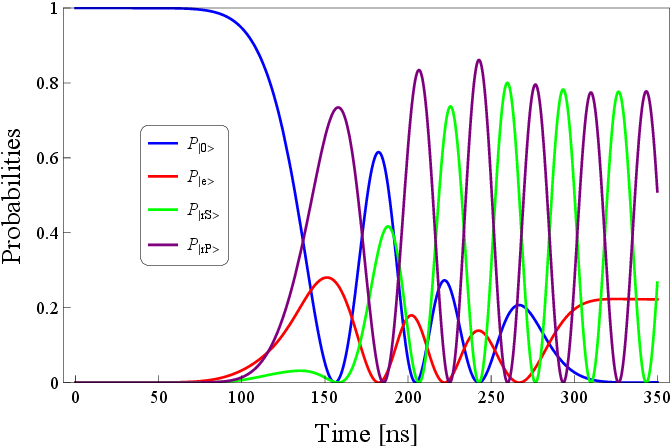} \\[3pt]
    \end{tabular}%
  }

\noindent\begin{minipage}{\textwidth}
    \captionsetup{justification=raggedright,singlelinecheck=false}
      \caption{
  Left: pulse shapes of microwave-dressed dipole-dipole gate acting on each of the ions.  
  Right: single ion populations vs time, without dipole-dipole interaction. The red line indicates the population of the intermediate state, whose population is unavoidable due to the presence of the micro-wave field, even when adiabacity is ensured. The parameters for the simulation are respectively: $\Omega_p/2\pi= \Omega_s/2\pi=$ 60 \text{MHz}, $\Omega_{mw}/2\pi=$ 30 \text{MHz}, $T=$ 70 \text{ns}, $\tau_p=$145 \text{ns}, $\tau_s=$205 \text{ns}. 
}
    \label{fig:2}
  \end{minipage}
\end{figure*}

We use the system of \cite{Zhang2020} as a model for our investigations. It consists of ${}^{88}$Sr${}^+$ ions confined in a linear Paul trap with an inter-ion separation of $4.2~\mu\text{m}$, whose level scheme is depicted in Fig.~\ref{fig:1}. The qubit is encoded in the Zeeman sublevels of the electronic ground and metastable states, with 
$\ket{1} = \ket{5S_{1/2}, m_J=-1/2}$ and $\ket{0} = \ket{4D_{5/2}, m_J=-5/2}$. 
The optical qubit transition is coherently driven by a narrow-linewidth laser at 674~nm, while projective state-dependent readout is performed by fluorescence detection on the $5S_{1/2} \leftrightarrow 5P_{1/2}$ transition using resonant light at 422~nm.  
The metastable state $\ket{0}$ is first coupled to the intermediate state $\ket{e} = \ket{6P_{3/2}, m_J=-3/2}$ using ultraviolet light at 243~nm, followed by coupling to the target Rydberg state $\ket{rS} = \ket{46S_{1/2}, m_J=-1/2}$ with a second ultraviolet beam at 306~nm and the quantization axis coincides with the trap axis. 
The effective two-photon Rabi frequencies are on the order of tens of megahertz, enabling fast population transfer.  
To generate strong interactions between two Rydberg-excited ions, the $\ket{rS}$ state is resonantly coupled to the nearby $\ket{rP} = \ket{46P_{1/2}, m_J=+1/2}$ level using a continuous-wave microwave field at 122~GHz. This microwave dressing produces hybrid eigenstates with induced dipole moments, resulting in a tunable dipole-dipole interaction between ions. 
The combined STIRAP and microwave dressing fields are depicted in the top left panel of Fig.~\ref{fig:2}. As we show below the microwave field acting simultaneously with STIRAP distorts the dark state and populates the intermediate state. In order to demonstrate this effect let us first introduce the three state Hamiltonian which reads,
\be
H_{3lvl}=\Delta\ket{e}\bra{e}+\frac{\Omega_p}{2}\ket{0}\bra{e}+\frac{\Omega_s}{2}\ket{e}\bra{rS}+h.c.
\ee

Transferring to the adiabatic basis with two consecutive Givens rotations around the first and last state with, 
\be
U=\left[
\begin{array}{ccc}
 \sin (\theta ) \sin (\phi ) & \cos (\phi ) & \cos (\theta ) \sin (\phi
   ) \\
 \cos (\theta ) & 0 & -\sin (\theta ) \\
 \sin (\theta ) \cos (\phi ) & -\sin (\phi ) & \cos (\theta ) \cos (\phi
   ) \\
\end{array}
\right],
\ee
where the angles read 
$\theta=\tan^{-1} \left(\Omega_s/\Omega_p\right),$ $\varphi=\frac{1}{2}\tan^{-1}\left(\Omega_{rms}/\Delta\right)$ and  $\Omega_{rms}=\sqrt{\Omega_s^2+\Omega_p^2}$
we end up with the following Hamiltonian in the adiabatic basis
\be\label{H3lvlSTIRAP}
H_{3Ad}= \lambda_+\ket{b_1}\bra{b_1}+\lambda_-\ket{b_2}\bra{b_2} + 0 \ket{d}\bra{d}, 
\ee
with $\lambda_{\pm}=\frac{1}{2}\Big(\Delta\pm\sqrt{\Delta^2+\Omega_{rms}^2} \Big)$ and the bright and dark states of the adiabatic basis defined as 
\bse
\begin{align}
\ket{b_1} &= \sin(\theta)\sin(\varphi)\ket{0}
           + \cos(\varphi)\ket{e}
           + \cos(\theta)\sin(\varphi)\ket{rS}, \label{eq:b1} \\[3pt]
\ket{b_2} &= \sin(\theta)\cos(\varphi)\ket{0}
           - \sin(\varphi)\ket{e}
           + \cos(\theta)\cos(\varphi)\ket{rS}, \label{eq:b2} \\[3pt]
\ket{d}   &= \cos(\theta)\ket{0}
           - \sin(\theta)\ket{rS}. \label{eq:d}
\end{align}
\ese

As we see from Eq.(\ref{H3lvlSTIRAP}) the dark state does not participate in the interaction and the $\ket{e}$ state remains unpopulated thus the system does not decay through it.

\paragraph*{} This picture changes rapidly with the addition of more states. Once we add a fourth state, representing the $\ket{rP}$ and a constant microwave field to this system the new Hamiltonian will read,
\be
\begin{split}
H_{4lvl}=&\frac{\Omega_p}{2}\ket{0}\bra{e}+\frac{\Omega_s}{2}\ket{e}\bra{rS}+\frac{\Omega_{mw}}{2}\ket{rS}\bra{rP} \\&+
\Delta\ket{e}\bra{e} + \Delta_{rr}\ket{rP}\bra{rP} +h.c.\label{H_{1Q}}
\end{split}
\ee
We can transform with the same $U,$ acting on the first three states, the four level Hamiltonian to
\be \label{H4lvlSTIRAP}
H_{4Ad}=\left[
\begin{array}{cccc}
 \lambda_+ & 0 & 0 &
   \frac{\Omega_{mw} \Omega_s \sin (\gamma ) }{2\Omega_{rms}} \\
 0 & 0 & 0 & -\frac{\Omega_{mw} \Omega_p}{2 \Omega_{rms}} \\
 0 & 0 &\lambda_- & \frac{\Omega_{mw}\Omega_s \cos (\gamma )}{2\Omega_{rms}} \\
 \frac{\Omega_{mw}\Omega_s \sin (\gamma )}{2 \Omega_{rms}} &
   -\frac{\Omega_{mw} \Omega _p}{2 \Omega_{rms}} & \frac{\Omega_{mw}
    \Omega_s \cos (\gamma )}{2 \Omega_{rms}} & \Delta_{rr} \\
\end{array}
\right],
\ee 
where $\gamma=\frac{1}{2}\cot^{-1}(\Delta/\Omega_{rms}).$


\begin{figure*}[t]
  \noindent\makebox[\textwidth]{%
    \begin{tabular}{cc}
    \includegraphics[width=0.52\textwidth]{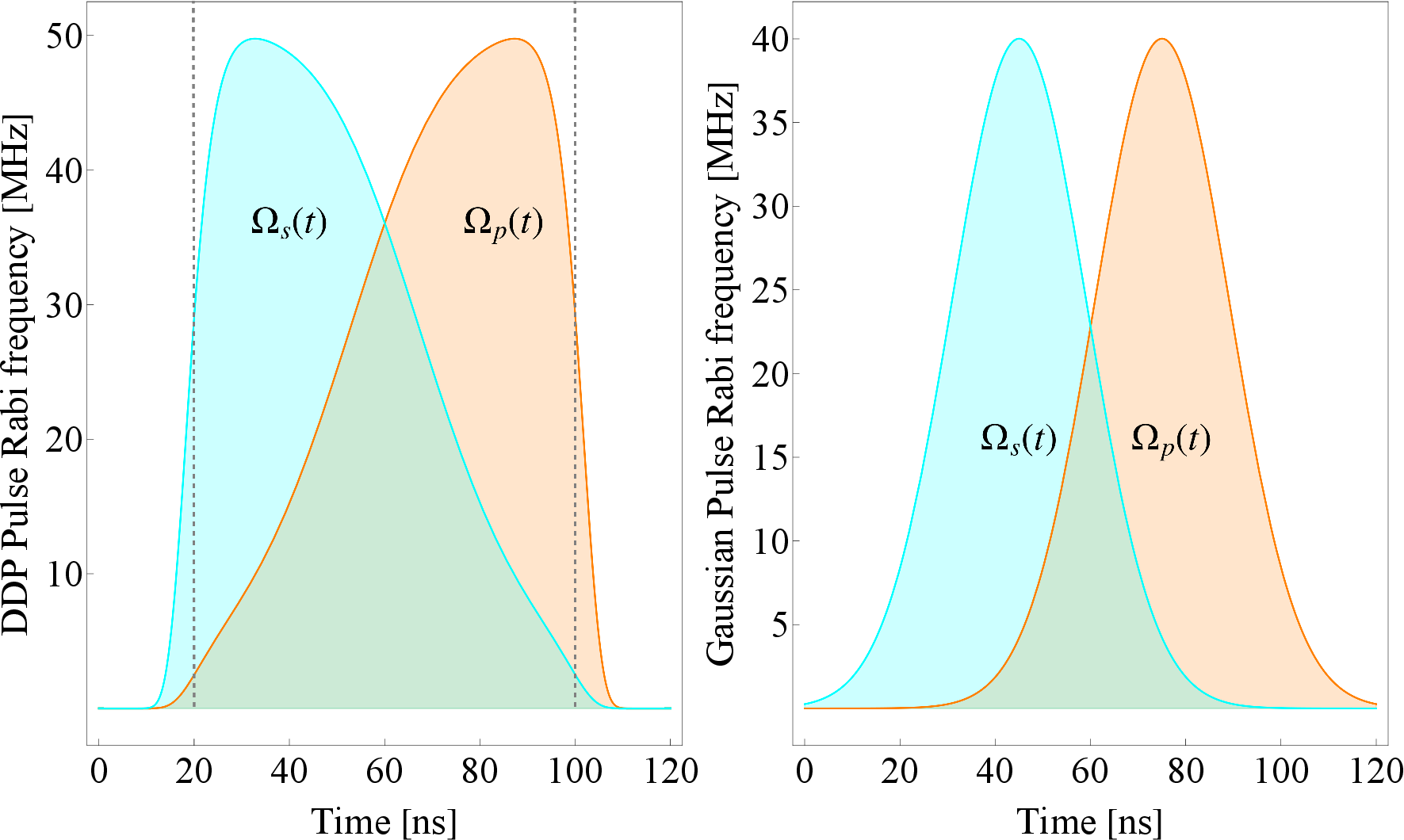} &
    \includegraphics[width=0.48\textwidth]{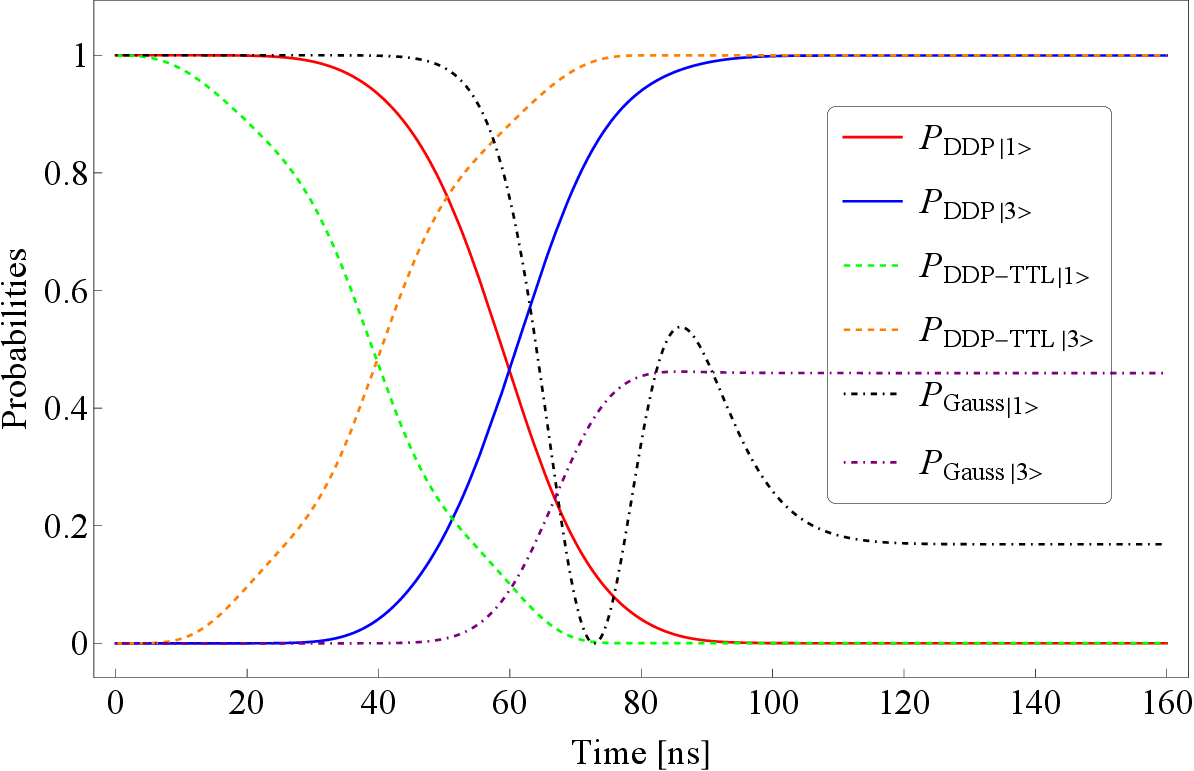} \\[5mm]
    \includegraphics[width=0.5\textwidth]{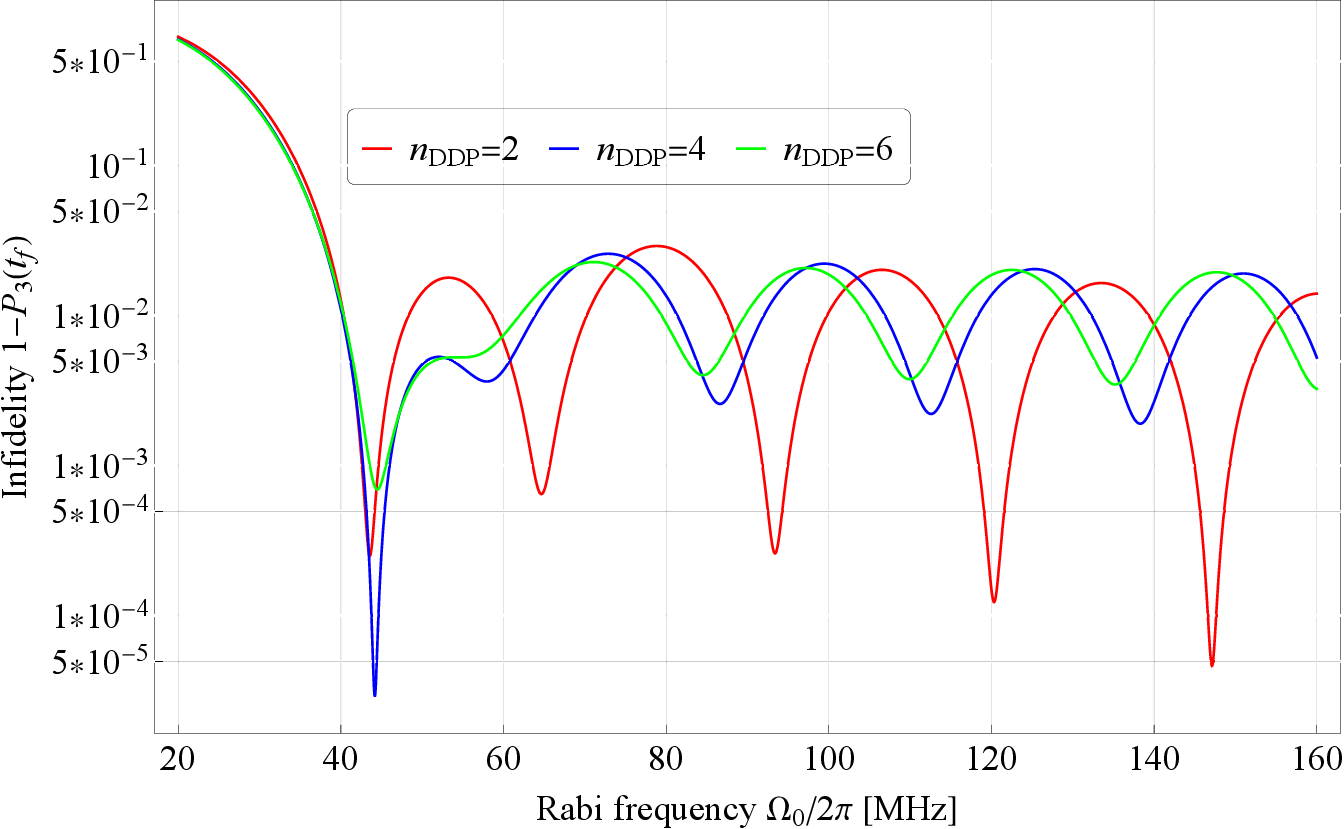} &
    \includegraphics[width=0.5\textwidth]{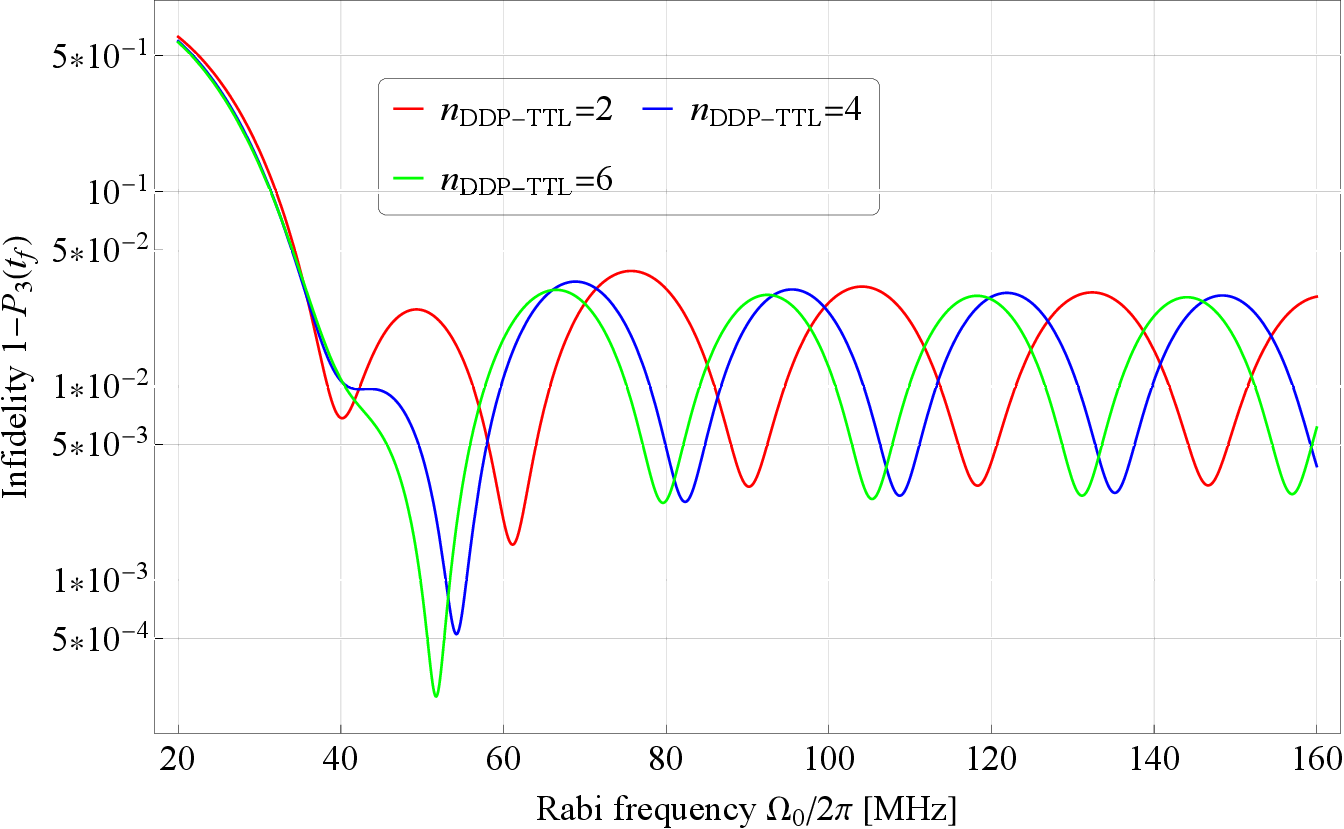}
  \end{tabular}%
  }

\noindent\begin{minipage}{\textwidth}
    \captionsetup{justification=raggedright,singlelinecheck=false}
      \caption{(Color online)
  Top: DDP-optimized(left) and Gaussian(center) pulse shapes used in the simulation of the population dynamics(right). The DDP pulses can further be shortened, indicated by the gray lines, with a TTL pulse. 
  Bottom: Infidelity of the STIRAP process with DDP and TTL shortened DDP pulses for different asymmetry parameter n. The parameters of the simulations are as follows: DDP$\to$ $\lbrace  \Delta/2\pi = 20$ MHz, $\Omega_0/2\pi =  44.07$, MHz,  
$T = 140$ ns, n = 4, $\alpha = 3$, and $\tau_{s}= \tau_{p} = 60$ ns. $\rbrace$, DDP-TTL $\to$ $\lbrace \Delta/2\pi = 15$ MHz, 
$\Omega_0/2\pi = 51.67$ MHz,  
$T = 140$ ns, $n = 6$, $\alpha = 3$, and $\tau_{s} =\tau_{p} = 40$ ns $\rbrace$, Gaussian Pulses $\to \lbrace \Delta/2\pi = 20$MHz, $\Omega_0/2\pi = 40$MHz, 
$T = 20$ns, $\tau_s = 45$ns and $\tau_p = 75$ns$\rbrace$.}  
    \label{fig:3}
  \end{minipage}
\end{figure*}
Based on the Hamiltonian of Eq.({\ref{H4lvlSTIRAP}) we now illustrate the problem arising form the microwave dressing of the Rydberg states in Fig.~\ref{fig:2}. We simulate the STIRAP excitation quite similar to that in~\cite{Zhang2020} with a few minor differences, so that we can demonstrate that the leakage channel is independent of the pulse shape. In our simulation we use Gaussian pulses (top left panel of Fig.~\ref{fig:2}) of the form,
\bse
\be
\Omega_p(t)= \Omega_p \exp\Big\{ -\frac{(t-\tau_p)^2}{T^2}\Big\} 
\ee
\be
\Omega_s(t)= \Omega_s \exp\Big\{-\frac{(t-\tau_s)^2}{T^2}\Big\},
\ee
\be
\Omega_{mw}(t)= \Omega_{mw},
\ee
\ese
while trigonometric functions were used in~\cite{Zhang2020}. Also the constant microwave field is an order of magnitude weaker (see all parameters in Fig.~\ref{fig:2}). 
As we can see from Eq.(\ref{H4lvlSTIRAP}), what used to be a dark state is now coupled to the $\ket{rP}$ state by the two-photon channel created by the pump and the microwave field. This leads to a significant population of the intermediate $\ket{e}$ state as illustrated in the top right panel of Fig.~\ref{fig:2}. 
 This is the prime reason for the population of the intermediate state (red curve) and the deviation from adiabacity is the secondary. The simultaneous action of the STIRAP stage and the dressing field thus results in a significant distortion of the bright-dark state picture, where population is scattered through all states. Yet this distortion does not deviate the gate substantially from the original proposal~\cite{Rao2014}. The motivation behind this setup is that for a two-qubit gate, where each qubit is under the influence of the STIRAP and microwave fields, and the qubits interact with dipole-dipole coupling, the entangling phase between them can be calculated as 
\be
\varphi(t)= V_R \int_{0}^{t} P_{rr}(t')\,\mathrm{d}t'.
\ee 
This is justified if the system evolves adiabatically in the weak interaction limit $V_R \ll \Omega_{P,S}$, also we can see from Eq.~\eqref{H4lvlSTIRAP}, that the coupling to the intermediate state is proportional to $\Omega_{rms}^{-1},$ thus it can be strongly suppressed with large Stokes field. Under these conditions the gate time for a maximally entangling $\pi-$phase gate is given as $\tau=\frac{8\pi}{3V_R}.$ There are three main challenges with this gate. First, the fidelity depends strongly on the adiabacity of the STIRAP and mostly to the coupling to the intermediate state. Although it can be suppressed, it can never be fully eliminated, which cannot result in a ultrahigh (
$>$99.99$\%$ ) fidelity. Second, the gate time can be lowered only to the extent to which the $V_R \ll \Omega_{P,S}$ regime of excitation is maintained. Also the process is strongly dependent on the STIRAP, for shorter gate, one needs unattainable ($\sim$THz) Rabi frequencies to compensate for the adiabacity of the process and to suppress the coupling to the intermediate state. Third, the original proposal~\cite{Rao2014}, was designed for a single Rydberg state and thus assumed that the population can be brought back at any time, depending only on the acquired phase. If the dipole-dipole interaction is induced, this is not the case. The population then, can only be brought back when it is in the $\ket{rS}$ state due to the selection rules for the second STIRAP. Thus only phases that are acquired during a round-trip to the starting point of the Rydberg manifold can be realized.

\section{Non-Adiabatic Speed-up}\label{sec:Three}
The gate we propose aims at $(i)$ reducing the gate time $\tau_g,$ $(ii)$ maximising the fidelity $F_{return}=|\langle \psi(t_i)| \psi(\tau_g)\rangle|^2$ and $(iii)$ allowing for arbitrary phase to be generated within the gate time. The way we achieve this is by separating the STIRAP and the dipole interaction stages of excitation. In this section we focus on the first step, by speeding up the STIRAP stage.

In order for STIRAP to work well, the adiabatic condition has to be well satisfied, its violation leads to population of the intermediate state and consequential decay. Aiming for gates in the nanosecond regime may require power much beyond the available. In the system under investigation the Rydberg states are excited by two-photon excitation using UV lasers. Since such laser sources are scarce usually these wavelengths are obtained by a process of wave-mixing, in some cases more than once, which further reduces the available laser power. Thus realistic Rabi frequencies are on the order of 50-60 MHz. In this case the adiabatic condition is barely satisfied and one has to resort either to counter-diabatic(CD) driving or pulse-shaping to overcome this issue. The former is also quite challenging in a ladder system, since the CD field will have to involve a direct coupling to the Rydberg state. This implies a wavelength that lies deep in the UV or X-ray spectrum. Therefore one has to rely on the latter approach which is pulse shaping in order to combat the non-adiabatic transitions.
A popular method for minimizing such transitions is based on the Dykhne, Davis and Pechukas (DDP) approximation.

In the context of STIRAP, the idea is to reduce the original three-level system to an effective two-level problem, either via single-photon resonance or in the large-detuning limit. In both cases, the optimization condition takes a simple form: $\Omega_{rms}$ must remain constant throughout the evolution. When this condition is satisfied, the DDP approximation predicts the suppression of transition points in the complex time plane, leading to vanishing nonadiabatic losses and, therefore, ultrahigh-fidelity transfer.

This was demonstrated in \cite{Vasilev2009} where to implement this condition in practice, the pump and Stokes fields are parameterized as
\begin{align} 
  \Omega_p(t) &= \Omega_0 F(t)\,\sin \left[\tfrac{\pi}{2} f(t)\right],\\
  \Omega_s(t) &= \Omega_0 F(t)\,\cos \left[\tfrac{\pi}{2} f(t)\right],\label{DDP_PS}
\end{align}
where $\Omega_0$ sets the peak scale of the couplings. 
Here, $f(t)$ is an auxiliary monotonic function interpolating smoothly between $f(-\infty)=0$ and $f(+\infty)=1$, which determines the effective mixing of the two fields. 
A common choice is a logistic form,
\begin{equation}
  f(t) = \frac{1}{1+e^{-\lambda t/T}},
\end{equation}
with $\lambda$ controlling the steepness and $T$ the characteristic timescale of the transition.

The envelope $F(t)$ acts as a ``mask'' that ensures finite pulse areas and defines the overall temporal width of the interaction. 
A hypergaussian form is typically used,
\begin{equation}
  F(t) = \exp \left[-\left(\tfrac{t}{T_0}\right)^{2n}\right],
\end{equation}
where $n=1$ corresponds to a Gaussian, and larger $n$ yield flatter plateaus in the overlap region of the pump and Stokes fields. 
This construction guarantees that the two fields overlap with nearly constant rms Rabi frequency while still vanishing smoothly at large times.

In this way, the DDP-optimized fields $\Omega_p(t)$ and $\Omega_s(t)$ realize adiabatic passage that is both highly robust and capable of achieving error probabilities below the fault-tolerance threshold for quantum information processing.

The dynamics and the infidelity of the pulses of Eqs.(\ref{DDP_PS}), compared to the standard STIRAP evolution is illustrated in Fig.~\ref{fig:3}. We assume that each pulse takes a bit more than 100 ns and each pulse shape is equally detuned on 20 MHz from the intermediate state.  We can see in the top right frame that on the same time scale of 120 ns, that is while the evolution is non-adiabatic, the standard STIRAP with Gaussian pulses fails, while the DDP-optimized pulses work quite well. Note that for the DDP pulses nothing happens in the first and last 20 ns. This allows us to further shorten the pulses by a TTL pulse, thus combining it with the AOM as an additional tool for the gate speed-up. The bottom frames show the infidelity of the full DDP pulses (left) and the TTL modified (right). Evidently where we cut the pulses and the asymmetry (the n value) plays a huge role, since for roughly the same power of  45 MHz (51 MHz for the DDP-TTL pulses) the DDP pulse of n=4 reaches infidelity lower than $10^{-5},$ while for the shorter pulses it is lower than $10^{-4}.$ Although we lost an order of magnitude in the infidelity we gained 40 ns reduction in gate time, without significant power overhead. 

%
\begin{figure*}[t]
  \noindent\makebox[\textwidth]{%
    \begin{tabular}{cc}
    \includegraphics[width=0.47\textwidth]{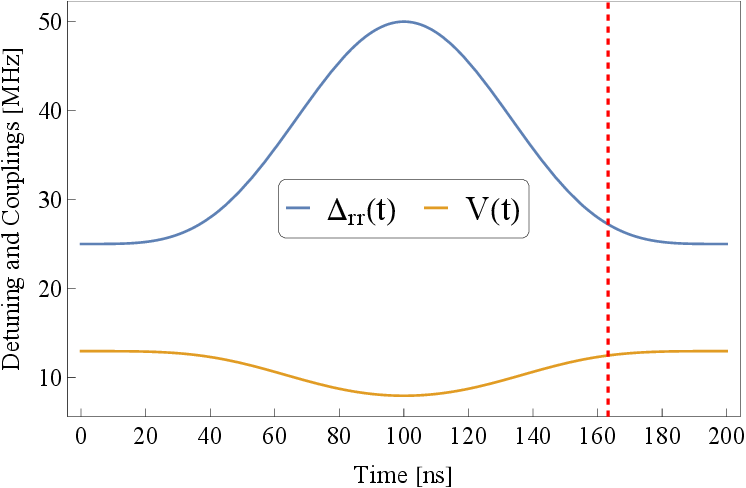} &
    \includegraphics[width=0.48\textwidth]{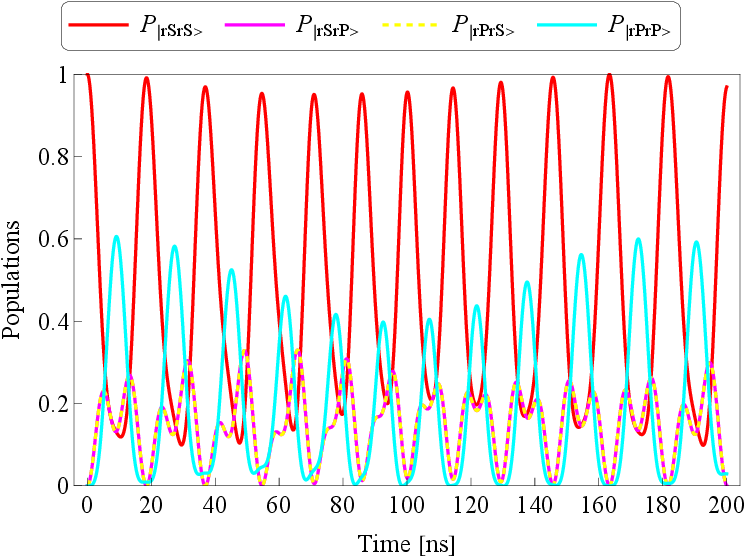} \\
    \includegraphics[width=0.48\textwidth]{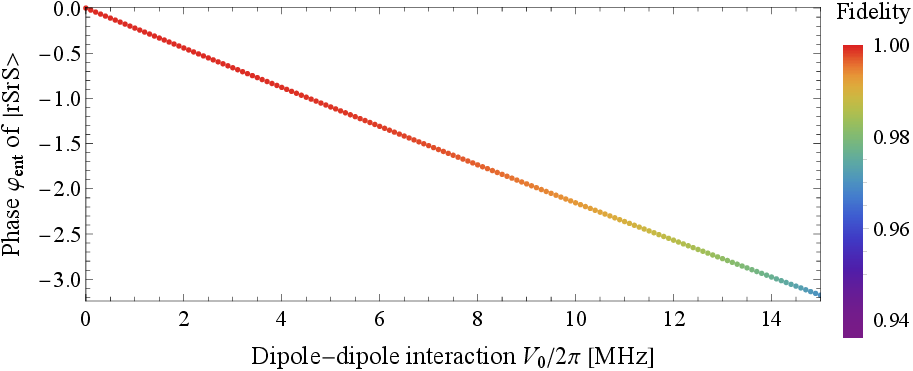} &
    \includegraphics[width=0.48\textwidth]{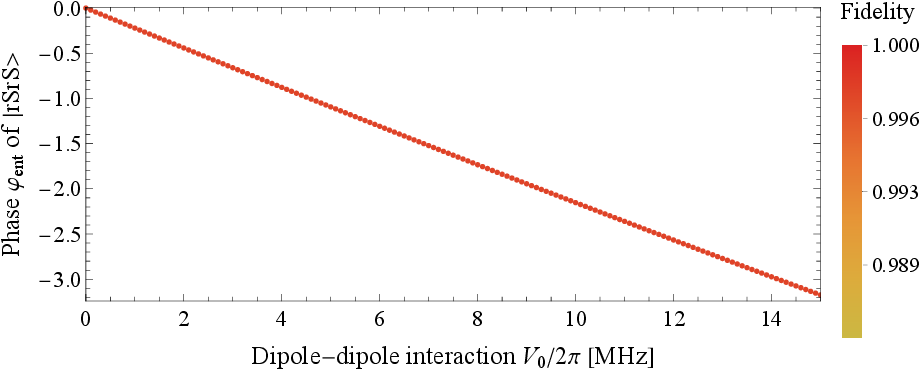}
  \end{tabular}%
  }

\noindent\begin{minipage}{\textwidth}
    \captionsetup{justification=raggedright,singlelinecheck=false}
      \caption{
   Top frames: time dependence of the $\Delta_{rr}(t)$ and $V(t)$ (a) and the induced population dynamics in the Rydberg manifold by constant Rabi (b). The dashed line in (a) indicates the gate time at which complete population return (CPR) occurs.
   Bottom frames: entangling phase of $\ket{rSrS}$ without (c) and with (d) gate time compensation. We note that $\varphi_{ent}$ is compensated with the constant offset, that is accumulated when $V_0=0.$
}
    \label{fig:4}
  \end{minipage}
\end{figure*}

\section{Two-qubit gate}\label{sec:Four}
In order to describe the full double STIRAP process, including the dipole-dipole interaction we follow the original model of \cite{Zhang2020}, where the Hamiltonian reads,

\begin{subequations}
\begin{equation}
H_I = V_0 \frac{\Omega_{\mathrm{mw}}(t)^2}{\Omega_{\mathrm{mw}}(t)^2 + \Delta_{rr}(t)^2}( \ket{rSrP}\bra{rPrS}+\ket{rPrS}\bra{rSrP}).
\label{eq:DD_Interaciton}
\end{equation}
\begin{equation}
H_{2Q} = \sum_{k} H_{1Q_k} + H_I ,
\end{equation}\label{eq:DDP_PS}
\end{subequations}
with $H_{1Q}$ given by Eq.~\eqref{H_{1Q}}.

Our modified interaction protocol however starts with STIRAP to the Rydberg states, $U_{DDP \uparrow}\ket{00}\to\ket{rSrS}$ followed by dipole-dipole interaction $U_{DD}\ket{rSrS} \to \exp(i \varphi_{ent})\ket{rSrS}$ and brought back to the computational basis via  $U_{DDP\downarrow}\ket{rSrS}\to \exp(i (\varphi_{ent}+\varphi_{loc}))\ket{00}.$ 
With perfect resonant STIRAP, following the dark state does not generate dynamical or geometric phase, but rather only gauge phase since $U_{STIRAP}\ket{c_1}\to-\ket{c_3}.$ Non-resonant STIRAP or violation of the adiabatic condition can account for additional phase accumulation. The local phase $\varphi_{loc}$ thus incorporates all such factors. Within an experiment it can be estimated and can be corrected or accounted for.
The estimation of the entangling phase $\varphi_{ent}$ is the problem we focus on, but before we examine it further let us track the evolution of the system first. Since the STIRAP stage and the dipole-dipole interaction stage are separated, it is worthwhile to investigate only the dynamics of the Rydberg manifold. In explicit form the two-qubit Hamiltonian then reads
\bse
\be
H_{2QRyd}=\frac{1}{2}\left(
\begin{array}{cccc}
 0 & \Omega _{\text{mw}} & \Omega _{\text{mw}} & 0 \\
 \Omega _{\text{mw}} & 2\Delta _{\text{rr}}(t) & 2V(t) & \Omega _{\text{mw}}\\
 \Omega _{\text{mw}} & 2V(t) & 2\Delta _{\text{rr}}(t) & \Omega _{\text{mw}} \\
 0 & \Omega _{\text{mw}} & \Omega _{\text{mw}} & 4 \Delta _{\text{rr}}(t) \\
\end{array}\label{DD4lvl}
\right)
\ee
\ese
acting only in the basis $\{\ket{rSrS},\ket{rSrP},\ket{rPrS},\ket{rPrP}\}.$


\begin{figure*}[t]
  \noindent\makebox[\textwidth]{%
    \begin{tabular}{cc}
    \includegraphics[width=0.49\textwidth]{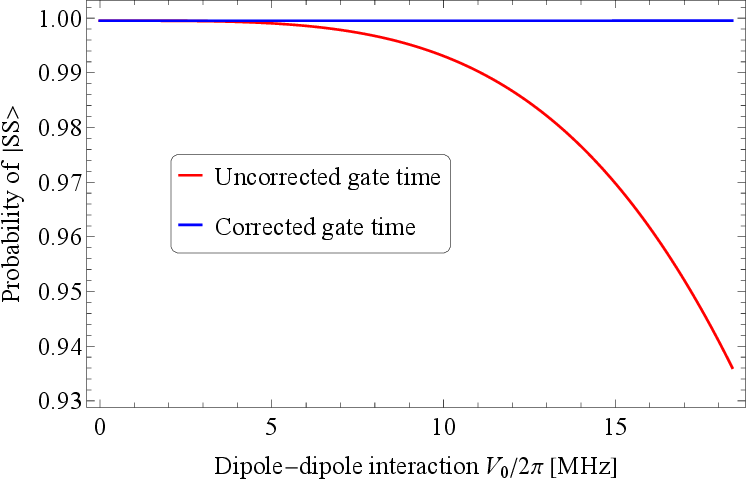} &
    \includegraphics[width=0.5\textwidth]{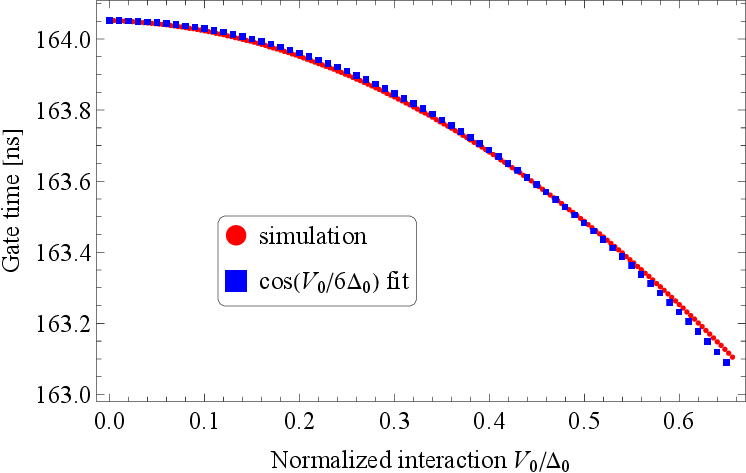} \\
    \includegraphics[width=0.51\textwidth]{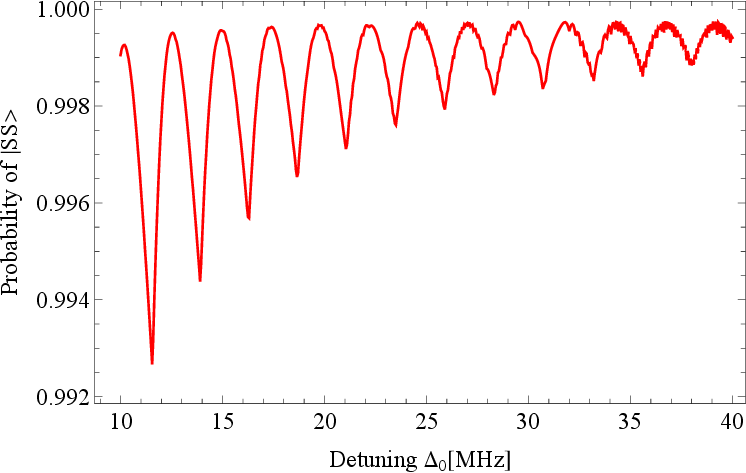} &
    \includegraphics[width=0.53\textwidth]{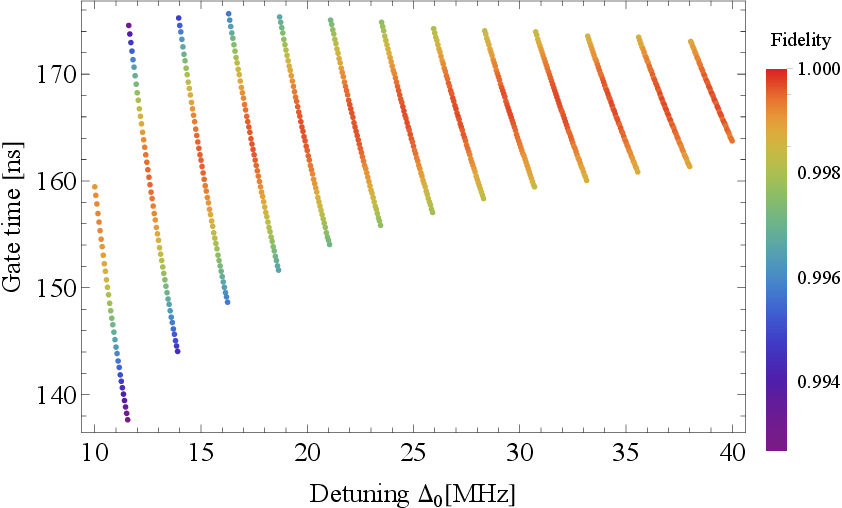}
  \end{tabular}%
  }

\noindent\begin{minipage}{\textwidth}
    \captionsetup{justification=raggedright,singlelinecheck=false}
      \caption{
   Top frames: probability for complete population return (CPR) in the $\ket{rSrS}$ state (left) with and without gate time compensation. Gate time drift for the normalized dipole-dipole interaction (right).
   Bottom frames: CPR of $\ket{rSrS}$ (left) and optimal gate time (right) as a function $\Delta_0.$  
}
    \label{fig:5}
  \end{minipage}
\end{figure*}

\subsubsection{Resonant microwave excitation}
As we see from Eq.~\eqref{eq:DD_Interaciton} the maximal strength of the dipole-dipole interaction is when the microwave field is resonant. This type of interaction is not optimal, which is best illustrated if the excitation is also constant besides resonant. The propagator can then be evaluated analytically by taking the exponent of the Hamiltonian $U_{DD}=\exp(i H_{2QRyd} t)_{|\Delta_{rr}=0}$.  In order to bring the population back to the computational states, the system must undergo CPR to the initial state $\ket{rSrS}$ and acquire the necessary entangling phase during the evolution in the Rydberg manifold. Then the only relevant propagator element is $U_{DD_{11}}$ which reads,
\be
\begin{split}
U_{DD_{11}}= \frac{1}{2} \left(1+e^{i\frac{V_0 t}{2} } \left(\cos \left(\frac{t}{2}  \sqrt{4 \Omega _{\text{mw}}^2+V_0^2}\right)\right.\right.\\\left.\left.-\frac{i V_0 \sin \left(\frac{t}{2}  \sqrt{4 \Omega _{\text{mw}}^2+V_0^2}\right)}{\sqrt{4 \Omega
   _{\text{mw}}^2+V_0^2}}\right)\right). \label{Ures11}
\end{split}
\ee
Further analysis of Eq.~\eqref{Ures11} shows us that, we can never pick up a $\pi$ phase since in order to do so the state must end at $-\ket{rSrS}.$ Due to the $1/2$ factor this is forbidden, because $Re\big(U_{DD_{11}}\big)=0,$ only when $Im\big(U_{DD_{11}}\big)=0.$ Further, in order to have CPR we need $|U_{DD_{11}}|^2=1$ at the end of the microwave excitation to return in the $\ket{0},\ket{1}$ basis. But this can only happen if a specific relations between $\Omega_{mw}, V_0$ and $t$ are fulfilled. Namely, $V_0 t = 2k\pi$ with $k>1$ and $\Omega_{mw}=\frac{|V_0|}{2}\sqrt{(c_0^2-1)}$ with the integer constant $c_0>1.$ In practice these constraints mean that we either pick a phase but we cannot achieve CPR, or we do not pick up phase and achieve it. 
\paragraph*{}The analytical solution for this model allows us to show that  resonant excitation, under separate dipole-dipole stage, cannot be used for the realization of a Rydberg two qubit phase gate. This was also noted in \cite{Giudici2025} where the authors were forced into overlapping it with the STIRAP stage, as in the original proposal of \cite{Rao2014}. This raises the question of whether these limitations can be mitigated through pulse shaping. As we argue below, the answer is negative. If we switch to a rotated basis by two Givens rotations over the i and j states $R(\pi/4)_{i,j},$ that is $\{\ket{rSrS},\ket{rSrP},\ket{rPrS},\ket{rPrP}\} \to \frac{1}{\sqrt{2}} \{\ket{rSrS}+\ket{rPrP},\ket{rSrP}+\ket{rPrS},\ket{rPrS}-\ket{rSrP},\ket{rPrP}-\ket{rSrS}\},$ then the transformed Hamiltonian reads
\be
\tilde{H}=\left(
\begin{array}{cccc}
 0 & \Omega _{\text{mw}}(t) & 0 & 0 \\
 \Omega _{\text{mw}}(t) & V_0 & 0 & 0 \\
 0 & 0 & -V_0 & 0 \\
 0 & 0 & 0 & 0 \\
\end{array}
\right).
\ee 
We see that here the dipole-dipole interaction plays the role of the detuning, and the problem can be reduced to a two level system with two dark states. The propagator will have the same block structure $\tilde{U}_{DD}=diag(U_{2lvl},\exp(i\varphi_1),\exp(i\varphi_2))$ with 
\be
U_{2lvl}= \left[
\begin{array}{cc}
 a & b \\
 -b^* & a^* \\
\end{array}
\right]
\ee
corresponding to the bright block and trivial phase factors ($\varphi_1,\varphi_2$) that account for the evolution of the dark states. Once we rotate back to the original basis the corresponding element for the $\ket{rSrS}$ state reads
\be
U_{DD_{11}}= \frac{1+a}{2},
\ee
which suffers from the same constraints as the constant excitation. Evidently pulse shaping does not help, therefore we have to turn to non-resonant time-varying excitation.

\subsubsection{Non-resonant microwave excitation}

\begin{figure*}[t]
  \noindent\makebox[\textwidth]{%
    \begin{tabular}{cc}
    \includegraphics[width=0.49\textwidth]{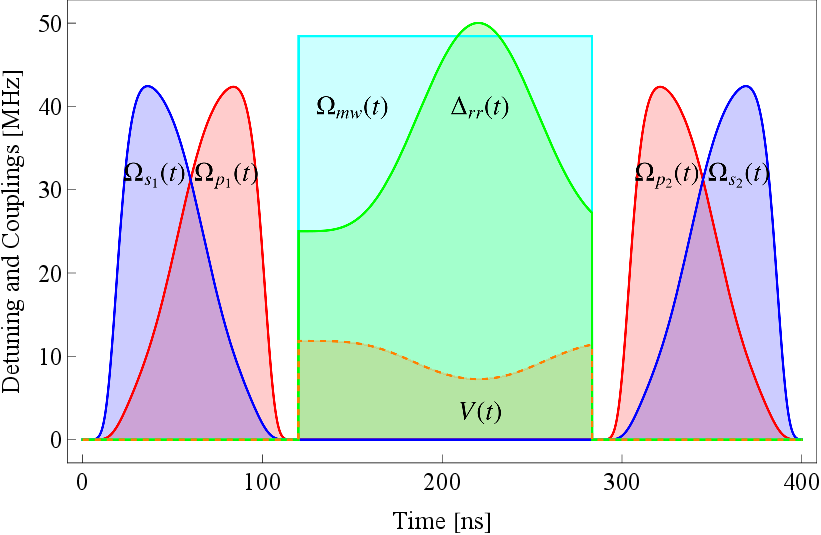} &
    \includegraphics[width=0.5\textwidth]{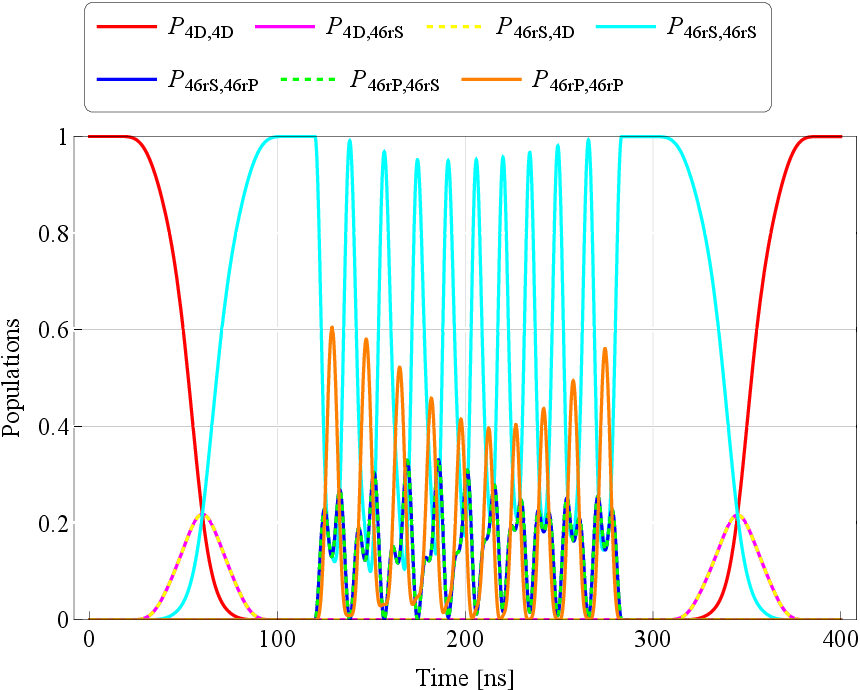} \\
  \end{tabular}%
  }

\noindent\begin{minipage}{\textwidth}
    \captionsetup{justification=raggedright,singlelinecheck=false}
      \caption{
   Couplings and detunings for the full two-qubit gate (left) and the induced dynamics (right). The parameters for the DDP-STIRAP are the same as in Fig.~\ref{fig:3}, for the dipole-dipole stage we have the field acting from 120 to 283.27 ns with $\Delta_0/2\pi=25$MHz, $\Omega_{mw}/2\pi=48.412$ MHz, $V_0=14.72$MHz, which corresponds to a phase of $\varphi_{ent}\approx\pi$ and the phase of the detuning function is $\phi=-3\pi/5.$ The fidelity for the full gate is $99.93\%.$
}
    \label{fig:6}
  \end{minipage}
\end{figure*}

Although we believe this system can be solved analytically, we were unable to derive such a solution, therefore we can only explore the system numerically. However the resonant excitation serves as a reference point, based on which we estimated that the most relevant quantities are the detuning and the dipole-dipole interaction. They need to be different and used for two distinct purposes. The detuning has to be employed in such a way that CPR can be achieved and with $V(t)$ we want to control the accumulated entangling phase. Physically this seems different from the standard implementation of \cite{Zhang2020}, where the interaction is fixed and the accumulated phase is controlled by the time of the interaction. However it does not differ much in its mechanism, since $V(t)$ is controlled by the distance between the ions, and the prefactor  $\frac{\Omega_{\mathrm{mw}}(t)^2}{\Omega_{\mathrm{mw}}(t)^2 + \Delta_{rr}(t)^2},$ whose time dependence is determined only by the detuning, for constant Rabi. The strength of the current approach is that for almost no experimental overhead, it allows a shorter gate time, it does not couple to the intermediate state and allows dipole-dipole interaction beyond the perturbative regime. The gate time is approximately the same for all implemented phases and if $V_0$ is insufficient for the accumulation of large enough phase, only the dipole-dipole stage of the gate needs to be extended. 

There are various detuning functions that can be employed, the one which we found to fulfill all the requirements reads,
\be
\Delta_{rr}(t)=\Delta_0\Big(1+\sin\left( \frac{t\pi}{T}+\phi \right)^4\Big).
\ee
The phase $\phi$ is such that within any time interval $\Delta_{rr}(t)$ will be as in Fig.~\ref{fig:4} (a). Now for a time interval $T,$  over which the detuning and the dipole-dipole interaction are symmetric functions we can realize CPR at some specific gate time roughly around $0.82T$ (we note however, that this is parameter dependent). This means that the detuning and the interaction have to be asymmetric (note the red dashed line in frame (a) indicating the gate time). The population dynamics is illustrated in Fig.~\ref{fig:4} (b). The CPR at the gate time of around 164 ns only occurs if the Rabi and the detuning are related by
\be
\Omega_{mw}=\frac{\sqrt{15}}{2}\Delta_0.
\ee
The way we derive this relation is semi-empirical. It is based also on the resonant solution where we derived a factor between $\Omega_{mw}$ and $V_0$ that is $\frac{1}{2}\sqrt{c_0^2-1}=\frac{\sqrt{15}}{2}$ for $c_0=4.$ In the resonant case this relation ensured CPR, based on Eq.~\eqref{Ures11}. If we instead calculate the same $U_{DD}$ propagator, but rather this time for $V_0=0,$ and constant $\Delta_{rr}=\Delta_0$ we arrive to a similar expression that indicates a relation between $\Omega_{mw}$ and $\Delta_{0}$. Thus intuitively we assumed that this relation might hold for time varying $\Delta_{rr}(t)$ and $V(t),$ since at least it has to hold between $\Omega_{mw}$ and $V_0$ in the simplest case. This expectation is corroborated by the numerical simulations.
 
Simply realizing this excitation is necessary but not sufficient condition for a two-qubit C-Phase gate. What also has to happen is phase accumulation based on the dipole-dipole interaction $V(t).$ Only when the above relation between the Rabi and the detuning holds, can the phase be accumulated by simply changing $V_0.$ This is illustrated in Fig.~\ref{fig:4} (c). Evidently the phase changes linearly only with respect to $V_0,$ with all other parameters held fixed. This proves that an entangling phase is being accumulated since it is only caused by the dipole-dipole interaction. Experimentally, this smooth variation of $V_0$ can be controlled by the ion distance.
The changing color of the dots indicates a drop in the fidelity, which can be corrected by adjusting the gate time as in Fig.~\ref{fig:4} (d). This slow drift of the gate time is also manifested in the return probability of $\ket{rSrS},$ shown in Fig.~\ref{fig:5} (a). This is the more important issue, since the overall fidelity across the STIRAP and the dipole-dipole stages will suffer if there is residual population in the Rydberg manifold, that is not being brought back by the second STIRAP. We can however compensate for it, since the drift can be approximated by $\tau_g \cos\left(\frac{V_0}{6\Delta_0}\right)$ as illustrated in Fig.~\ref{fig:5} (b). Although the approximation is not perfect, experimentally it can be adjusted further. Based on the fit we can conclude that this drift is proportional to the $V_0/\Delta_0$ ratio. 
The choice of the detuning magnitude is bound by two factors, first we want it to be high enough so that we avoid the resonant excitation regime and second we need it low enough so that we can achieve a meaningful speed up of the gate. The peaks in Fig.~\ref{fig:5} (c) indicate that not every value is appropriate, especially once we determine the achievable interaction $V_0$ and the gate time. This is shown in frame (d). Changing $\Delta_0$ also changes the population dynamics, therefore periodically (approximately each 2.4 MHz) the gate time jumps a few tens of ns. For that reason we carried all simulations at $\Delta_0=25$ MHz, since it is a fair compromise between gate speed (of around 164 ns) and fidelity. Finally we show the full gate in Fig.~\ref{fig:6}, with the left frame illustrating the detuning  and couplings in all stages, and the right one showing the dynamics of the process. The overall fidelity we reach for return to the $\ket{00}$ state is of $99.93\%$ with entangling phase of $\varphi_{ent}\approx\pi.$

\section{Conclusion and Outlook}\label{sec:Five}
In this work we show that the Rydberg gates that employ addressing with STIRAP and microwave-dressing simultaneously have detrimental effects on the gate fidelity, due to the inevitable population of the intermediate state. We proposed a novel approach that separates the STIRAP stage and the dipole-dipole interaction. We demonstrated a speed up of the STIRAP stage by employing asymmetric pulse shapes based on the DDP approximation, which allowed us to bring a single STIRAP process to around 120 ns at experimentally feasible Rabi frequencies of about 44 MHz. We further demonstrated that induced dipole-dipole interaction by resonant excitation is inconvenient since either the gate fidelity will deteriorate, or no phase will be accumulated by the interaction. Since pulse shaping alone did not help either we demonstrated that a gate can only be achieved if the detuning is varied during the excitation in a non-trivial manner. An asymmetric shape is needed on top of specific parameter relations between $\Delta_0$ and $\Omega_{mw}$ so that the gate time will be approximately fixed, with predictable drift. Under such conditions the acquired entangling phase is a linear function of the magnitude of the interaction $V_0.$ Combining all stages of the interaction we were able to achieve a gate of 400 ns with fidelity of $99.93\%.$ This gate mechanism allows a dramatic speed up in atomic systems where the dipole-dipole interaction can reach 100 MHz and above. In Rydberg ions this remains the main limiting factor for the speed of the gate, however the main advantage of our scheme is that it allows the gate to be realized beyond the perturbative dipole-dipole interaction regime. This problem outlines the necessity for analytical solutions for four level systems such as the one from Eq.~\eqref{DD4lvl}. Such a solution can provide further improvements in robustness, fidelity and shorter gate time. 

\acknowledgements
 This research is supported by the Bulgarian national plan for recovery and resilience, Contract No. BG-RRP-2.004-0008-C01 (SUMMIT), Project No. 3.1.4 and by the European Union’s Horizon Europe research and innovation program under Grant Agreement No. 101046968 (BRISQ).

\end{document}